\documentclass[preprint,superscriptaddress,showpacs,aps]{revtex4}

\usepackage{graphicx}
\usepackage{subfigure}
\usepackage{amsmath}
\usepackage{amssymb}

\begin{document}

\title{The coordinate transformation and the  exact solutions of the Schr\"{o}dinger equation
with position-dependent effective mass}

\author{JU Guo-Xing}
\email{jugx@nju.edu.cn}
\affiliation{Department of Physics, Nanjing University,
Nanjing 210093, China}

\author{CAI Chang-Ying}
\affiliation{Department of Physics, Nanjing University,
Nanjing 210093, China}
\affiliation{Department of Physics, Jinggangshan University, Jian 343009,
China}

\author{XIANG Yang}
\affiliation{Department of Physics, Nanjing University,
Nanjing 210093, China}

\author{REN Zhong-Zhou}
\affiliation{Department of Physics, Nanjing University,
Nanjing 210093, China}
\affiliation{Center of Theoretical Nuclear Physics, National Laboratory
of Heavy-Ion Accelerator, Lanzhou 730000, China}

\begin{abstract}
Using the coordinate transformation method, we solve the one-dimensional Schr\"{o}dinger equation
with position-dependent mass(PDM). The explicit expressions for the potentials, energy eigenvalues and
eigenfunctions of the systems are given.
The eigenfunctions can be expressed in terms of the Jacobi, Hemite and generalized
Laguerre polynomials. All potentials for these solvable systems have an extra term $V_m$
which produced from the dependence
of mass on the coordinate, compared with that for the systems of constant mass. The properties of $V_m$
for several mass functions are discussed.
\end{abstract}


\maketitle

\section{Introduction}
In recent years, the study of quantum systems with
position-dependent  mass (PDM)  has become one of the active subjects
\cite{von,ben,Levl,dek1,dek2,pla,Mil,des,gon1,gon2,Alh,roy,koc,que1,que2,que3,que4,ou,yu,cai}.
These systems  have wide applications in condensed matter physics,
nuclear physics, semiconductor theory and other related fields. In the theoretical researches,
 many methods developed in the study of systems with constant mass
have been generalized to
the systems with PDM and a number of interesting results
have been produced. Of those researches, the searching of exact solutions of the quantum wave equations
(Schrodinger equation, Klein-Gordan equation and Dirac equation)
for these systems is
an important aspect.
The factorization method\cite{inf}, operator
methods \cite{old}, coordinate transformation
methods\cite{man,dab,lev,dutt}, supersymmetric quantum mechanics
\cite{wit,dab,Coo}, Lie Algebra approach\cite{wu,wu2,lev2},path integral approach\cite{kle}
 and so on that exactly solve the quantum
wave equations with constant mass
have been extended to  the systems with PDM.

In the present work, we will use the coordinate transformation
method to the PDM Schr\"odinger equations and obtain their solutions  for
several  potentials.
There are two kinds of coordinate transformations.
The first one connects two different solvable potentials such as the Coulomb potential and harmonic oscillator,
Rosen-Morse and generlized Scarf potentials\cite{dutt}.
The second one transforms the Schr\"odinger equation
into the second order differential equation which has solutions of the special
functions such as hypergeometric functions and confluent hypergeometric functions,
and so  may provide  systematical studies
to the exact solutions of the Schr\"odinger equation.
When combined with the shape invariance, the coordinate transformation method
provides an important way to
classify the solvable potentials\cite{lev,dutt,Coo}. We will discuss the coordinate transformation
method in the second sense in this paper.

The paper is organized as follows. In section \ref{ctpdm}, the coordinate transformation
will be introduced and exact
solutions of the one-dimensional Schr\"{o}dinger equation with PDM
 for several potentials will be given.
The eigenfunctions of these systems can be expressed in terms of Jacobi polynomials, Hemite polynomials
and generalized Laguerre  polynomials. In section \ref{mfp}, the effective potentials  produced
from the dependence of mass on the position of the particle
 are given and their properties are discussed for several mass functions. In the last section,
some remarks and discussions will be made.

\section{coordinate transformation,
PDM Schr\"{o}dinger equation  and their solutions}
\label{ctpdm}

When the mass of a particle depends on its position, the mass
and momentum operator no longer commute, so there are
several ways to define the kinetic energy operator  of the quantum system \cite{von}. In this paper, we will
adopt the form of the kinetic energy introduced by BenDaniel and
Duke\cite{ben}, and the Hamiltonian with the position-dependent
effective mass $M(\vec{\bf r})$ and potential energy $V(\vec{\bf r})$
reads
\begin{equation}\label{ham}
  H=\vec{\bf P}\cdot\frac{1}{2M(\vec{\bf r})}\vec{\bf P}+V(\vec{\bf r})
          =-\frac{\hbar^{2}}{2m_{0}}\biggl[\vec{\bf \nabla}\cdot\frac{1}
          {m(\vec{\bf r})}\vec{\bf\nabla}\biggr]+V(\vec{\bf r}),
\end{equation}
where $m_{0}$ is a constant mass and $m(\vec{\bf r})$ is a
dimensionless position-dependent mass. Using natural units
$(m_{0}=\hbar=1)$ and only considering one-dimensional system, we get the
following Hamiltonian  from Eq. (\ref{ham})
\begin{equation}\label{ham-1}
H=-\frac{1}{2}\left[\frac{d}{dx}\frac{1}
          {m(x)}\frac{d}{dx}\right]+V(x)
          =-\frac{1}{2m}\frac{d^{2}}{dx^{2}}+\frac{m'}{2m^{2}}\frac{d}{dx}+V(x),
\end{equation}
where $m'$ denotes  the first derivative of $m(x)$ with respect to the coordinate $x$.
The PDM Schr\"{o}dinger equation corresponding to the Hamiltonian (\ref{ham-1}) reads
\begin{equation}
\label{eq2}
\frac{d^2\psi (x)}{dx^2} - \frac{m'}{m}\frac{d\psi
(x)}{dx} + 2m\left[ {E - V (x)} \right]\psi (x) = 0.
\end{equation}
Now, we make the following coordinate transformation to the eigenfunction $\psi(x)$ in  Eq. (\ref{eq2})
\cite{man,dab,lev,dutt}
\begin{equation}
\label{eq4}
\psi (x) = f(x)F(g(x)),
\end{equation}
where $F(g)$ is some special function and satisfies the following second order differential
equation
\begin{equation}
\label{eq5}
\frac{d^2F}{dg^2} + Q(g)\frac{dF}{dg} + R(g)F(g) =0.
\end{equation}
$Q(g)$ and $R(g)$ in Eq. (\ref{eq5})
are determined by the types of  special functions.
In this paper, we will consider three types of  special functions.
Substituting Eq. (\ref{eq4}) into Eq. (\ref{eq2}), one gets the equation
\begin{equation}
\label{eq6} \frac{d^2F}{dg^2} + \left[ \frac{g''}{(g')^2} +
\frac{2f'}{fg'} - \frac{m'}{mg'} \right]\frac{dF}{dg} + \left[
\frac{f''}{f(g')^2} - \frac{m'f'}{mf(g')^2} + \frac{2m\left[ E -
V(x) \right]}{(g')^{2}} \right]F = 0.
\end{equation}
On comparing Eq. (\ref{eq5}) with Eq. (\ref{eq6}), we may have the following relations
\begin{equation}
\label{eq7} Q(g) = \frac{g''}{(g')^2} + \frac{2f'}{fg'} -
\frac{m'}{mg'},
\end{equation}
\begin{equation}
\label{eq8} R(g) = \frac{f''}{f(g')^2} - \frac{m'f'}{mf(g')^2} +
\frac{2m\left[ {E - V(x)} \right]}{(g')^2}.
\end{equation}
From Eqs. (\ref{eq7}) and (\ref{eq8}), we have
\begin{equation}
\label{eq9} \frac{f'}{f} = \frac{1}{2}\left( {Qg' -
\frac{g''}{g'} + \frac{m'}{m}} \right),
\end{equation}
\begin{equation}
\label{eq12}
E - V(x) = \frac{(g')^2}{2m}\left[ {R(g) -
\frac{1}{2}\frac{dQ}{dg} - \frac{1}{4}Q^2} \right] +
\frac{1}{4m}\left[ {G(g') - G(m)} \right],
\end{equation}
where $G(z) = \frac{z''}{z} -
\frac{3}{2}\left({\frac{z'}{z}}\right)^2$.
On making integration to Eq. (\ref{eq9}), we  get the transformation
function $f(x)$ to be of the form
\begin{equation}\label{eq13}
f(x) = \sqrt {\frac{m}{g'}} \exp \left({\frac{1}{2}\int^{g(x)} {Q(g)dg} }
\right),
\end{equation}
which has an extra factor $\sqrt{m}$ compared with that for the constant mass\cite{lev}.

Now, it follows that the PDM Schr\"odinger equation can be exactly solved if the forms of functions
$Q(g)$ and $R(g)$ (or the types of the special functions $F(g)$) are given for a mass $m(x)$ that is a function
of space coordinate $x$, and if the right hand side of Eq. (\ref{eq12})
can be divided into two parts: one depends on the quantum number $n$, but is independent of $x$, while another part
depends on $x$ rather than the quantum number $n$. For the above division, the former
is the eigenvalue $E$ of the system and the latter is  the corresponding potential $V(x)$.
Such system is exactly solvable  with eigenfunction $\psi(x)$, energy eigenvalue $E$ and  potential $V(x)$.

In the following sections, we will take $F(g)$
to be of the Jacobi, Hermite and generalized Laguerre polynomials, respectively.
And for the convenience, we introduce the auxiliary function
\begin{equation}
\label{eq14}
\mu (x) = \int^{x} {\sqrt {m(x)} dx.}
\end{equation}
$G(g')$ in Eq.(\ref{eq12}) can be rewritten into the following form in terms of $\mu(x)$
\begin{equation}\label{gmeq}
    G(g')=G(\mu')+(\mu')^2G(\frac{dg}{d\mu}),
\end{equation}
where
\begin{equation}\label{gmeq1}
    G(\frac{dg}{d\mu})=\frac{d^3g}{d\mu^3}/\left(\frac{dg}{d\mu}\right)-\frac{3}{2}\left[
    \frac{d^2g}{d\mu^2}/\left(\frac{dg}{d\mu}\right)\right]^2,
\end{equation}
which reduces to $G(g')$ for the constant mass.
Eqs. (\ref{eq12}) and (\ref{gmeq}) show that there are a new term $\frac{1}{4m}(G(\mu')-G(m))$
and an extra factor $\frac{1}{m}$ in each term compared with that for the constant
system\cite{lev}. When $m$ depends on the coordinate $x$, the term $\frac{1}{4m}(G(\mu')-G(m))$
 will be grouped into
the potential $V(x)$ of the system. In this sense, the PDM induces effective interaction between particles
in the system, which
is consistent with the concept of effective mass in the condensed matter physics and other related fields. We will
denote $V_m$ in the following sections  as
\begin{equation}\label{ptm1}
    V_m=\frac{1}{4m}(G(m)-G(\mu'))=\frac{1}{8m}\left[\frac{m''}{m}-\frac{7}{4}\left(\frac{m'}{m}\right)^2\right],
\end{equation}
which is attributed  to the dependence of the mass $m$ on $x$.

\subsection{Jacobi polynomial and solvable potentials}

When we choose $F(g)$ to be the  Jacobi polynomial $ P_n^{(\alpha,\beta)}
(g)$,  the corresponding differential equation for  $ P_n^{(\alpha,\beta)}
(g)$ gives the expressions of $Q(g), R(g)$ in Eq. (\ref{eq5})\cite{mag,and}
\begin{subequations}
\label{jp12}
\begin{equation}
\label{jp12a}
Q(g) = \frac{{ - \alpha + \beta}}{{1 - g^2 }} - \frac{(2 + \alpha +
\beta)g}{1 - g^2 },
\end{equation}
\begin{equation}
\label{jp12b}
R(g) = \frac{{n(1 + \alpha + \beta + n)}}{{1 - g^2 }},
\end{equation}
\end{subequations}
where $\alpha$ and $\beta$ are parameters, $n=0,1,2,\cdots$. Substituting Eq. (\ref{jp12})
into Eq. (\ref{eq12}), we get
\begin{equation}
\label{jp14}
\begin{aligned}
 E - V(x) =& \frac{1}{4m}[G(g')-G(m)] + n(1 + \alpha + \beta + n)\frac{{g'^2 }}{2m(1 - g^2 )}
  \\
  &+\left[2(2+\alpha+\beta)- (\alpha - \beta)^2\right] \frac{{g'^2 }}{8m(1 - g^2 )^2 } \\
           & + (\beta + \alpha)(\beta - \alpha)\frac{{gg'^2 }}{{4m(1 - g^2 )^2 }}
  + \left[1-(1 + \alpha + \beta)^2\right] \frac{{g^2 g'^2 }}{{8m(1 - g^2 )^2 }}.
 \end{aligned}
\end{equation}
Now we will chose appropriate $g(x)$ to make RHS of Eq. (\ref{jp14}) have a term that is independent
of $x$ , but may contain $n$. Once $g(x)$ is determined, then  we will get
solvable potentials $V(x)$
and its corresponding energy eigenvalue $E$ from Eq. (\ref{jp14}). Inserting $g(x)$ into Eq. (\ref{eq13}),
we will obtain the explicit expression of
$f(x)$, and so the eigenfunction of the system is given by $\psi (x) = f(x)P_n^{(a,b)}(g)$
for the  above potential $V(x)$.
Similar to those for the systems with constant mass, there are two cases that
satisfy the above requirements, and  each has several different functions $g(x)$.

{\bf Case 1:} $g(x)$ satisfies the differential equation
\begin{equation}\label{geq1}
  \frac{g'^2}{(1 - g^2 )m(x)} =C,
\end{equation}
where $C$ is a constant. In this case, one has
\begin{equation}\label{gg1}
   G(\frac{dg}{d\mu})=-C\left[1+\frac{3g^2}{2(1-g^2)}\right].
\end{equation}
Each solution $g(x)$ to Eq. (\ref{geq1})  corresponds to an exactly solvable system.
According to the procedures given above, we can get its potential, eigenfunction and energy eigenvalues.
We will not give the details of calculations and  just list the results.
Note that the energy eigenvalues are chosen such that $E_{n=0}=0$. The definitions of the parameters $s$, $\lambda$
are made so that the potential, eigenfunctions and energy eigenvalues can reduce to those for the system with
constant mass when $m(x)=1$, respectively\cite{dab,lev}.

(i) $g(x) = i\sinh(a\mu(x))$, $C=-a^2$
\begin{subequations}
\begin{equation}
\label{e31}
 E_n  = \frac{1}{2}s^2a^2-\frac{1}{2}a^2(s-n)^2,
\end{equation}
\begin{equation}
\label{e32}
\begin{aligned}
V(x)=&  \frac{1}{2}s^2a^2+ \frac{1}{2}a^2 ( \lambda^2-s^2-s)
 {\rm sech}^{2}(a\mu(x))  \\
  &-\frac{1}{2}a^2\lambda(2s+1)\tanh(a\mu(x)){\rm sech}(a\mu(x))+V_{m},
\end{aligned}
\end{equation}
\begin{equation}
\label{e33}
\begin{aligned}
 \psi (x) =& \left[ {m(x)}\right]^{\frac{1}{4}}\left(\cosh(a\mu)\right)^{-s}
 \exp \left\{ (-\lambda\tan^{-1}[\sinh (a\mu(x))]\right\}
\\
&\times P_n^{(-i\lambda-s-\frac{1}{2}, i\lambda-s -\frac{1}{2})} (i\sinh (a\mu(x))),
 \end{aligned}
\end{equation}
\end{subequations}
where $s=-\frac{1}{2}(\alpha+\beta+1),\lambda=-\frac{1}{2}i(\beta-\alpha)$.

(ii) $g(x) = \cosh(a\mu(x))$, $C=-a^2$
\begin{subequations}
\begin{equation}
\label{e34}
 E_n  = \frac{1}{2}s^2a^2-\frac{1}{2}a^2(s-n)^2,
\end{equation}
\begin{equation}
\label{e35}
\begin{aligned}
V(x) =&  \frac{1}{2}s^2a^2+\frac{1}{2}a^2(\lambda^2+s^2+s){\rm cosech}^2(a\mu(x))
 \\
 &- a^2\lambda\left(s+\frac{1}{2}\right) \coth(a\mu(x)){\rm cosech}(a\mu(x))+ V_m,
\end{aligned}
\end{equation}
\begin{equation}
\label{e36}
\begin{aligned}
 \psi (x) =& \left[ {m(x)}\right]^{\frac{1}{4}}(\sinh(a\mu(x)))^{-s}
 \left(\frac{1-\cosh(a\mu(x))}{1+\cosh(a\mu(x))}\right)^{\frac{\lambda}{2}}
  \\
 &\times P_n^{(\lambda-s-\frac{1}{2}, -\lambda-s -\frac{1}{2})} (i\sinh (a\mu(x))),
  \end{aligned}
\end{equation}
\end{subequations}
where $s=-\frac{1}{2}(\alpha+\beta+1),\lambda=-\frac{1}{2}(\beta-\alpha)$.

(iii) $g(x) = \cos(a\mu(x))$, $C=a^2$
\begin{subequations}
\begin{equation}
\label{e37}
 E_n  = -\frac{1}{2}s^2a^2+\frac{1}{2}a^2(s+n)^2,
\end{equation}
\begin{equation}
\label{e38}
 \begin{aligned}
 V(x) =&  -\frac{1}{2}s^2a^2 +\frac{1}{2}a^2(\lambda^2+s^2-s)\csc^2(a\mu(x))
 \\
  &- \frac{1}{2}a^2\lambda(2s-1)\cot(a\mu(x))\csc(a\mu(x))
  +V_{m},
\end{aligned}
\end{equation}
\begin{equation}
\label{e39}
 \begin{aligned}
 \psi (x) = \left[ {m(x)}\right]^{\frac{1}{4}}(\sinh(a\mu(x)))^{s}
 \left(\frac{1+\cos(a\mu(x))}{1-\cos(a\mu(x))}\right)^{\frac{\lambda}{2}}  
   P_n^{(-\lambda+s -\frac{1}{2}, \lambda+s -\frac{1}{2})}(\cos(a\mu(x)), 
 \end{aligned}
\end{equation}
\end{subequations}
where $s=\frac{1}{2}(\alpha+\beta+1),\lambda=\frac{1}{2}(\beta-\alpha)$.

(iv) $g(x) = \sin [a\mu(x)]$, $C=a^2$
\begin{subequations}
\begin{equation}
\label{e40}
 E_n  = -\frac{1}{2}s^2a^2+\frac{1}{2}a^2(s+n)^2,
\end{equation}
\begin{equation}
\label{e41}
\begin{aligned}
V(x)=&-\frac{1}{2}s^2a^2+ \frac{1}{2}a^2(\lambda^2+s^2-s)\sec^2(a\mu(x))
 \\
  &-\frac{1}{2}\lambda a^2(2s-1)\sec(a\mu(x))\tan(a\mu(x)) +V_m,
\end{aligned}
\end{equation}
\begin{equation}
\label{e42}
\begin{aligned}
 \psi(x) = \left[ {m(x)}\right]^{\frac{1}{4}}\left(\cos(a\mu(x))\right)^{s}
 \left(\frac{1+\sin(a\mu(x))}{1-\sin(a\mu(x))}\right)^{\frac{\lambda}{2}}
  P_n^{(-\lambda+s-\frac{1}{2},\lambda+s -\frac{1}{2})}(\sin(a\mu(x))),
 \end{aligned}
\end{equation}
\end{subequations}
where $s=\frac{1}{2}(\alpha+\beta+1),\lambda=\frac{1}{2}(\beta-\alpha)$.

{\bf Case 2:} $g(x)$ satisfies the differential equation
\begin{equation}\label{geq2}
  \frac{g'^2}{(1 - g^2 )^2m(x)} =C,
\end{equation}
where $C$ is a constant. Now, we have
\begin{equation}\label{gg2}
   G(\frac{dg}{d\mu})=-2C.
\end{equation}
We will consider four solutions of $g$ to Eq. (\ref{geq2}). The corresponding potential, eigenfunction and
energy eigenvalue of the exactly solvable system  for each $g$ are listed as follows:

(i) $ g(x) =\tanh (a\mu(x))$, $C=a^2$
\begin{subequations}
\begin{equation}
\label{e44}
E_n  =  \frac{1}{2}s^2a^2+\frac{\lambda^2}{2s^2}a^2- \frac{1}{2}a^2 \left[(s -n)^2  + \frac{{\lambda ^2
}}{{(s - n)^2 }}\right],
\end{equation}
\begin{equation}
\label{e45}
\begin{aligned}
 V(x)& =  \frac{1}{2}s^2a^2+\frac{\lambda^2}{2s^2}a^2 - \frac{1}{2}a^2s(s + 1){\rm sech}^2(a\mu(x))
 - \lambda  a^2 \tanh(a\mu(x))  +V_m,
 \end{aligned}
\end{equation}
\begin{equation}
\label{e46}
\begin{aligned}
 \psi (x) =\left[ {m(x)}\right]^{\frac{1}{4}}(\cosh(a\mu(x)))^{s+n}
 \left(\frac{1-\tanh(a\mu(x))}{1+\tanh(a\mu(x))}\right)^{\frac{\bar{a}}{2}}
  P_n^{(s - n - \bar{a},s - n + \bar{a})} (\tanh (a\mu(x))) ,
 \end{aligned}
\end{equation}
\end{subequations}
where
\begin{equation}
\label{e47}
 \alpha = s - n + \bar{a},\;\; 
 \beta = s - n -\bar{a},\;\; 
\bar{a}=\frac{\lambda}{s-n}.
\end{equation}

(ii) $ g(x) =\coth(a\mu(x))$, $C=a^2$
\begin{subequations}
\begin{equation}
\label{e48a}
 E_n  = \frac{1}{2}a^2s^2+\frac{\lambda^2}{2s^2}a^2-\frac{1}{2}a^2\left[(s+n)^2+
 \frac{\lambda^2}{(s+n)^2}\right],
\end{equation}
\begin{equation}
\label{e49a}
\begin{aligned}
 V(x) = \frac{1}{2}a^2s^2+\frac{\lambda^2}{2s^2}a^2
 +\frac{1}{2}a^2 s(s-1)\csc^2(a\mu(x)) -\lambda a^2\cot(a\mu(x))+V_m,
 \end{aligned}
\end{equation}
\begin{equation}
\label{e50a}
\begin{aligned}
\psi (x) =& \left[ {m(x)}\right]^{\frac{1}{4}}\left(\sinh(a\mu(x))\right)^{n+s}
\left(\frac{\coth(a\mu(x))-1}{\coth(a\mu(x))+1}\right)^{\frac{\bar{a}}{2}}
 \\
 & \times P_n^{(-s - n - \tilde a, -s - n + \tilde a)} ( \coth(a\mu(x))),
\end{aligned}
\end{equation}
\end{subequations}
where
\begin{equation}
\label{e50b}
 \alpha = -s - n + \bar{a},\;\;
 \beta = -s - n -\bar{a},\;\;
\bar{a}=\frac{\lambda}{s+n}.
\end{equation}

(iii) $ g(x) =  - i\cot(a\mu(x))$, $C=-a^2$
\begin{subequations}
\begin{equation}
\label{e48}
 E_n  = -\frac{1}{2}a^2s^2+\frac{\lambda^2}{2s^2}a^2+\frac{1}{2}a^2(s-n)^2-
 \frac{1}{2}a^2\frac{\lambda^2}{(s-n)^2},
\end{equation}
\begin{equation}
\label{e49}
\begin{aligned}
 V(x) = -\frac{1}{2}a^2s^2+\frac{\lambda^2}{2s^2}a^2
 +\frac{1}{2}a^2 s(s+1)\csc^2(a\mu(x)) -\lambda a^2\cot(a\mu(x))+V_m,
 \end{aligned}
\end{equation}
\begin{equation}
\label{e50}
\begin{aligned}
\psi (x) =\left[ {m(x)}\right]^{\frac{1}{4}}\left(\sin(a\mu(x))\right)^{n-s}
 \exp [ a\bar{a}\mu(x)] 
  P_n^{(s - n +i\bar{a},s - n -i\bar{a})} ( - i\cot(a\mu(x))),
\end{aligned}
\end{equation}
\end{subequations}
where
\begin{equation}
\label{e50c}
 \alpha = s - n +i\bar{a},\;\;
 \beta = s - n -i\bar{a},\;\;
 \bar{a}=\frac{\lambda}{s-n}.
\end{equation}

(iv) $ g(x) =  - i\tan(a\mu(x))$, $C=-a^2$
\begin{subequations}
\begin{equation}
\label{e48d}
 E_n  = -\frac{1}{2}a^2s^2+\frac{\lambda^2}{2s^2}a^2+\frac{1}{2}a^2(s-n)^2-
 \frac{1}{2}a^2\frac{\lambda^2}{(s-n)^2},
\end{equation}
\begin{equation}
\label{e49d}
\begin{aligned}
 V(x) = -\frac{1}{2}a^2s^2+\frac{\lambda^2}{2s^2}a^2
 +\frac{1}{2}a^2 s(s+1)\sec^2(a\mu(x)) -\lambda a^2\tan(a\mu(x))+V_m,
 \end{aligned}
\end{equation}
\begin{equation}
\label{e504}
 \begin{aligned}
\psi (x) =\left[ {m(x)}\right]^{\frac{1}{4}}\left(\cos(a\mu(x))\right)^{n-s}
 \exp [-a\bar{a}\mu(x)] 
  P_n^{(s - n +i\bar{a},s - n -i\bar{a})} ( - i\tan(a\mu(x))),
\end{aligned}
\end{equation}
\end{subequations}
where
\begin{equation}
\label{e50d}
 \alpha = s - n +i\bar{a},\;\;
 \beta = s - n -i\bar{a},\;\;
 \bar{a}=\frac{\lambda}{s-n}.
\end{equation}

It is obvious that all above results reduce to those for the systems with constant mass when $m(x) = 1$
and $\mu(x)=x$\cite{dab,lev}. From  above explicit expressions for the potentials, eigenfunctions and energy eigenvalues
of the solvable systems, we see that the energy eigenvalues are the same as those for the systems with constant mass,
but eigenfunctions and potentials do not so when the mass of the particle depends on $x$. The effects of PDM
to eigenfunctions and potentials are twofold: the argument of
function $\mu(x)$ and an extra  factor containing $m(x)$ in the  eigenfunctions or a new term $V_m$ in the potentials.
In Section \ref{mfp}, we will
discuss the properties of $V_m$ for several mass functions $m(x)$. The above facts show that
PDM will make the classes of solvable potentials more general than those for the constant mass.

\subsection{Hermite polynomial and solvable potentials}

When $F(g)$ in Eq. (\ref{eq5}) is the Hermite polynomial, i.e. $F(g)=H_n(g)$,
then $Q(g)$ and $R(g)$ in Eq. (\ref{eq5}) have the following forms\cite{mag,and}, respectively
\begin{equation}
\label{eq16}
 Q(g) = - 2g,
 \quad
R(g) = 2n.\;\;\;(n = 0,1,2,\cdots)
\end{equation}
With above $Q(g)$, $R(g)$ and Eq. (\ref{eq12}), one has
\begin{equation}
\label{eq18}
E - V(x) = \frac{(g')^2}{2m}\left( {2n + 1 -
g^2} \right) + \frac{1}{4m}\left[ {G(g') - G(m)} \right].
\end{equation}

There are two cases of conditions with which
we can obtain the potentials and energy eigenvalues of the solvable systems
from Eq. (\ref{eq18}).

{\bf Case 1}: If $g(x)$ satisfies the equation
\begin{equation}
\label{eq19}
\frac{(g')^2}{m} = \omega,
\end{equation}
with $\omega > 0$, we  chose
\begin{equation}
\label{eq20}
g(x) = \sqrt \omega \mu (x).
\end{equation}
Inserting Eq. (\ref{eq20}) into Eq. (\ref{eq18}), we get the eigenvalue and potential of the system, respectively
\begin{subequations}
\begin{equation}
\label{eq21}
E_n = n\omega, 
\end{equation}
\begin{equation}
\label{eq22}
V(x) = -\frac{1}{2}\omega+\frac{1}{2}\omega ^2\left[ {\mu (x)}
\right]^2 + V_m.
\end{equation}
\end{subequations}
From Eqs. (\ref{eq16}), (\ref{eq20}) and (\ref{eq13}), the transformation function reads
\begin{equation}
\label{eq23} f(x) = \left[ {m(x)} \right]^{\frac{1}{4}}\exp ( { -
\frac{1}{2}g^2} ),
\end{equation}
so the corresponding eigenfunction of the system with potential (\ref{eq22}) is
\begin{equation}
\label{eq24}
\psi_n (x) = \left[ {m(x)}
\right]^{\frac{1}{4}}\exp ( - \frac{1}{2}g^2)H_n (g).
\end{equation}
It is obvious that (\ref{eq21}) is the same  as that  for the harmonic oscillator up to a constant term. When $m(x) =
1$,  Eqs. (\ref{eq22}) and (\ref{eq24}) reduce to the potential function
and eigenfunction of the harmonic oscillator, respectively.

{\bf Case 2}: If $g(x)$ is the solution of the differential equation
\begin{equation}
\label{eq25}
\frac{(g')^2}{m}g^2 = \frac{4\omega ^2}{(2n + 1)^2},
\end{equation}
where $\omega > 0$,  we have
\begin{equation}
\label{eq26}
g(x) = \sqrt {\frac{4\omega }{2n + 1}} \left[ {\mu
(x)} \right]^{\frac{1}{2}}.
\end{equation}
Putting Eq. (\ref{eq26}) into Eq. (\ref{eq18}),  we get
\begin{subequations}
\begin{equation}
\label{eq27}
E_n = 2\omega ^2- \frac{2\omega ^2}{(2n + 1)^2},
\end{equation}
\begin{equation}
\label{eq28}
 V(x) =2\omega ^2 - \frac{\omega }{2}\left[ {\mu (x)}
\right]^{ - 1} - \frac{3}{32}\left[ {\mu (x)} \right]^{ - 2} +V_m.
\end{equation}
\end{subequations}
Substituting Eq. (\ref{eq26}) into Eq. (\ref{eq13}), one has the transformation function of the form
\begin{equation}
\label{eq29}
f(x) = \left[ {m(x)\mu (x)}
\right]^{\frac{1}{4}}\exp ( { - \frac{1}{2}g^2} ),
\end{equation}
and the corresponding eigenfunction of the system reads
\begin{equation}
\label{eq30}
\psi _n (x) = \left[ {m(x)\mu (x)}
\right]^{\frac{1}{4}}\exp ( { - \frac{1}{2}g^2} )H_n (g).
\end{equation}

\subsection{Generalized Laguerre polynomial and solvable potentials}

When $F(g)$ in Eq. (\ref{eq5}) is the generalized Laguerre polynomial, i.e. $F(g)=L_n^{\alpha}(g)$,
$Q(g)$, $R(g)$ in Eq. (\ref{eq5}) will take the forms\cite{mag,and}, respectively
\begin{equation}
\label{eq32} Q(g) = \frac{\alpha+1}{g} - 1,
\quad
R(g) = \frac{n}{g}.\;\;\;(n =0,1,2,\cdots,\alpha
\ne - 1, - 2, -3,\cdots)
\end{equation}
Substituting Eq. (\ref{eq32}) into Eq. (\ref{eq12}), we obtain the relation
\begin{equation}
\label{eq34}
\begin{aligned}
E - V(x) = &\frac{(g')^2}{4mg}\left( {2n +
\alpha+1 } \right)+ \frac{(g')^2}{2mg^2}\left[ {\frac{(\alpha+1
)}{2} - \frac{(\alpha+1)^2}{4}} \right] \\
&- \frac{(g')^2}{8m}
+\frac{1}{4m}\left[ {G(g') - G(m)} \right].
\end{aligned}
\end{equation}

{\bf Case 1:}  When $g(x)$ satisfies the equation
\begin{equation}
\label{eq35} \frac{(g')^2}{mg} = 4\omega ,
\end{equation}
with $\omega > 0$,  we have
\begin{equation}
\label{eq36} g(x) = \omega \left[ {\mu (x)} \right]^2.
\end{equation}
Inserting Eq. (\ref{eq36}) into Eq. (\ref{eq34}), we get the energy eigenvalues and potential
of the system
\begin{subequations}
\begin{equation}
\label{eq40}
 E_n = 2n\omega,
\end{equation}
\begin{equation}
\label{eq41}
 V(x) = -\left(l + \frac{3}{2}\right)\omega+\frac{1}{2}\omega ^2\left[ {\mu (x)}
\right]^2 +\frac{1}{2}\frac{ l (l + 1)}{\left[ {\mu (x)} \right]^{ 2}} +
V_m,
\end{equation}
\end{subequations}
where
\begin{equation}
\label{eq39}
l=\alpha- \frac{1}{2}.\;\;\; {\left( {l
\ne - \frac{3}{2}, - \frac{5}{2}, - \frac{7}{2},\cdots} \right)}
\end{equation}
In this case, the transformation function is
\begin{equation}
\label{eq42}
 f(x) = \left[ {m(x)}
\right]^{\frac{1}{4}}[\mu(x)]^{l + 1}\exp (-\frac{1}{2}\omega\mu(x)^2).
\end{equation}
With Eqs. (\ref{eq4}) and (\ref{eq42}), we obtain the corresponding eigenfunction of the system
\begin{equation}
\label{eq43}
\psi _n (x) = \left[ {m(x)}
\right]^{\frac{1}{4}}[\mu(x)]^{l+ 1}\exp (-\frac{1}{2}\omega\mu(x)^2)
L_{n}^{(l+\frac{1}{2})}(\omega\mu(x)^2).
\end{equation}
If $l = 0,1,2,\cdots,$ and $l$ is
viewed as the angular momentum quantum number, then Eq. (\ref{eq40}) is the
energy eigenvalues for the three-dimensional harmonic oscillator\cite{fyuu}. When $m(x)=
1$, Eqs. (\ref{eq41}) and (\ref{eq43})
reduce to the potential and eigenfunction for the three-dimensional isotropic
harmonic oscillator, respectively.

{\bf Case 2:} If $g(x)$ satisfies the equation
\begin{equation}
\label{eq44}
\frac{(g')^2}{mg^2} = a^2,
\end{equation}
with $a \ne 0$, we choose
\begin{equation}
\label{eq45}
g(x) = \exp [ - a\mu (x)].
\end{equation}
With Eqs. (\ref{eq44}), (\ref{eq45}) and (\ref{eq34}), we obtain
the energy eigenvalues, potential of the system
\begin{subequations}
\begin{equation}
\label{eq48}
E_n =\frac{1}{2}a^2s^2- \frac{1}{2}a^2\left(s - n\right)^2,
\end{equation}
\begin{equation}
\label{eq49}
 V(x) = \frac{1}{2}a^2s^2+\frac{a^2}{8}\exp \left[ { - 2
a\mu (x)} \right] -\frac{1}{4}a^2\exp \left[ - a\mu (x) \right]
+V_m,
\end{equation}
\end{subequations}
where
\begin{equation}
    s = n + \frac{1 }{2}\alpha.
 \;\;\;(s \ne 0, \pm\frac{1}{2}, \cdots, \pm\frac{n}{2},\cdots)
\end{equation}
From Eqs. (\ref{eq32}), (\ref{eq45}), (\ref{eq4}) and (\ref{eq13}),
we have the eigenfunction of the system
\begin{equation}
\label{eq51}
\psi _n (x) = \left[ m(x)\right]^{\frac{1}{4}}\exp\left[(n-s)a\mu(x)\right]
\exp \left[ - \frac{1}{2}e^{a\mu(x)} \right] L_{n}^{(2s-2n)}(\exp(-a\mu(x))).
\end{equation}
It is follows that the expression (\ref{eq48}) is the energy eigenvalue for
 the Morse potential with zero angular momentum.
When $m(x)=1$ and with the appropriate parameters $a$, $b$ and $c$,
then (\ref{eq49}) and (\ref{eq51}) are Morse potential and
eigenfunction with zero angular momentum, respectively.

{\bf Case 3:} If $g(x)$ is the solution to the differential equation
\begin{equation}
\label{eq52}
\frac{(g')^2}{m} = 4\omega ^2,
\end{equation}
where $\omega > 0$, we take
\begin{equation}
\label{eq53}
g(x) = 2\omega \mu (x).
\end{equation}
If we make the replacements $\omega= \frac{a}{n +
l+1}$,\;$\alpha=2l+1$ ($l \ne - 1, -
\frac{3}{2}, - \frac{4}{2},\cdots$) and use Eqs. (\ref{eq53}) and (\ref{eq34}), we
have
\begin{subequations}
\begin{equation}
\label{eq56}
E_n = \frac{a^2}{2(l + 1)^2}- \frac{a^2}{2(n +l + 1)^2},
\end{equation}
\begin{equation}
\label{eq57}
V(x) =  \frac{a^2}{2( l+ 1)^2}- \frac{a}{\mu (x)} +
\frac{l (l + 1)}{2\mu (x)^ 2} + V_m.
\end{equation}
\end{subequations}
In this case, the transformation function can be written as
\begin{equation}
\label{eq58}
f(x) = \left[ {m(x)} \right]^{\frac{1}{4}}(\mu(x))^{l +
1}\exp \left[ - \frac{a}{n+l+1}\mu(x)\right],
\end{equation}
so the eigenfunction of the system  reads as
\begin{equation}
\label{eq59}
\psi _n (x) =  \left[ {m(x)} \right]^{\frac{1}{4}}(\mu(x))^{l +
1}\exp \left[ - \frac{a}{n+l+1}\mu(x)\right]L_{n}^{(2l+1)}(\frac{2a}{n+l+1}\mu(x)).
\end{equation}
If $a = Z(Z$ is the charge numbers of the particle), $l = 0,1,2,\cdots$,
and $l$ is regarded as the angular momentum quantum number,
then Eq. (\ref{eq56}) is just the energy eigenvalues for the three-dimensional
Coulomb potential\cite{fyuu}.
When $m(x)=1$, Eqs. (\ref{eq57}) and (\ref{eq59}) reduce to the three-dimensional Coulomb potential
and its eigenfunction, respectively.

\section{Mass functions and potentials}
\label{mfp}

In this section, we discuss the effective potentials $V_m$ and their properties
due to the dependence of mass on the coordinate
for several mass functions.

{\bf Example 1:}\;\; We take the effective mass function to be of the form
\begin{equation}\label{mf1}
m(x) = \left( {\frac{b + x^2}{1 + x^2}} \right)^2,
\end{equation}
which has been used in many studies\cite{pla,Alh,roy}.
Substituting Eq. (\ref{mf1}) into Eq. (\ref{eq13}), we get
\begin{equation}
\label{eq64} \mu(x) = x + (b - 1)\arctan x.
\end{equation}
With Eq. (\ref{ptm1}), the contribution to the potential from mass function is
\begin{equation}\label{mfv1}
    V_m=\frac{(b - 1)[3x^4 + 2(2 -b)x^2 - b]}{2(b +x^2)^4}.
\end{equation}
It is seen that $V_m=0$ when $b=1$, which corresponds to the system of constant mass.

When $0<b<1$ or $1<b<4$, $V_m$
has three extreme points
\begin{equation}\label{ep1}
    x=0,\;\;\;
    x=\pm\left(b-1+\sqrt{\frac{1}{3}(2b^2-2b+3)}\right)^{\frac{1}{2}}.
\end{equation}
If $b>4$, then there are five extreme points for $V_m$, three of which  has the same form as that in
Eq. (\ref{ep1}), the other two extreme points are
\begin{equation}\label{ep2}
    x=\pm\left(b-1-\sqrt{\frac{1}{3}(2b^2-2b+3)}\right)^{\frac{1}{2}}.
\end{equation}
The characteristic curves for $V_m$  with different values of the parameter $b$ are depicted in Fig.1.
It is seen that  $V_m$ behaves like a barrier and it decrease as the parameter approaches to 1 from $b<1$.
While when $b>1$, $V_m$ just looks like a well. In this sense, $V_m$ will bound the motion of the particle.

\begin{figure}[htb]
\subfigure[]
{
\label{fig:mini:subfig:a}
\begin{minipage}[c]{0.3\textwidth}
\centering\includegraphics[width=5.0cm,height=5.5cm]{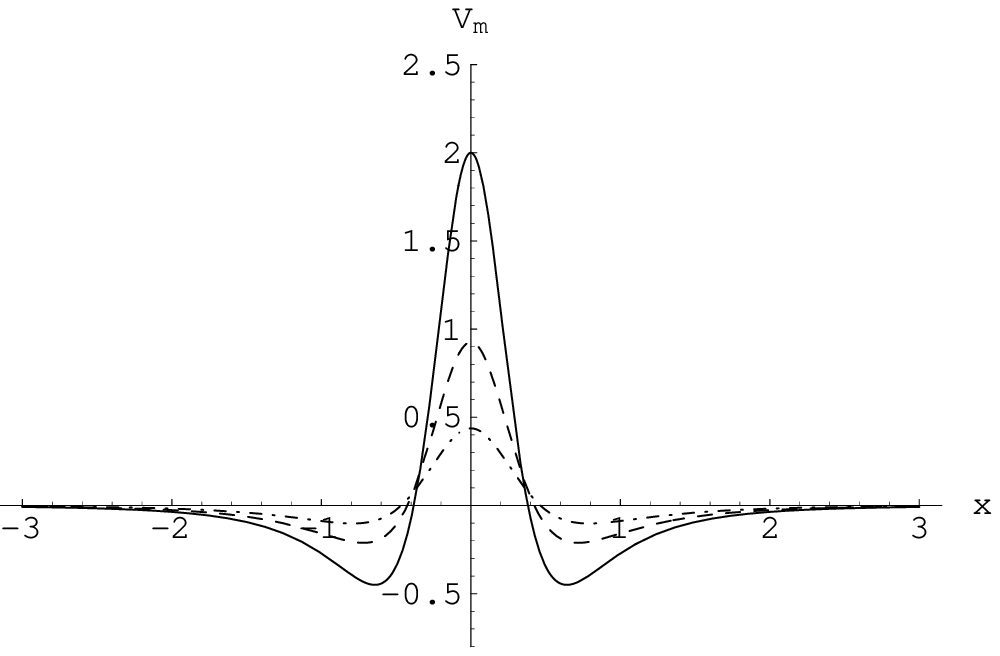}
\end{minipage}}%
\subfigure[]
{
\label{fig:mini:subfig:b}
\begin{minipage}[c]{0.3\textwidth}
\includegraphics[width=5.0cm,height=5.5cm]{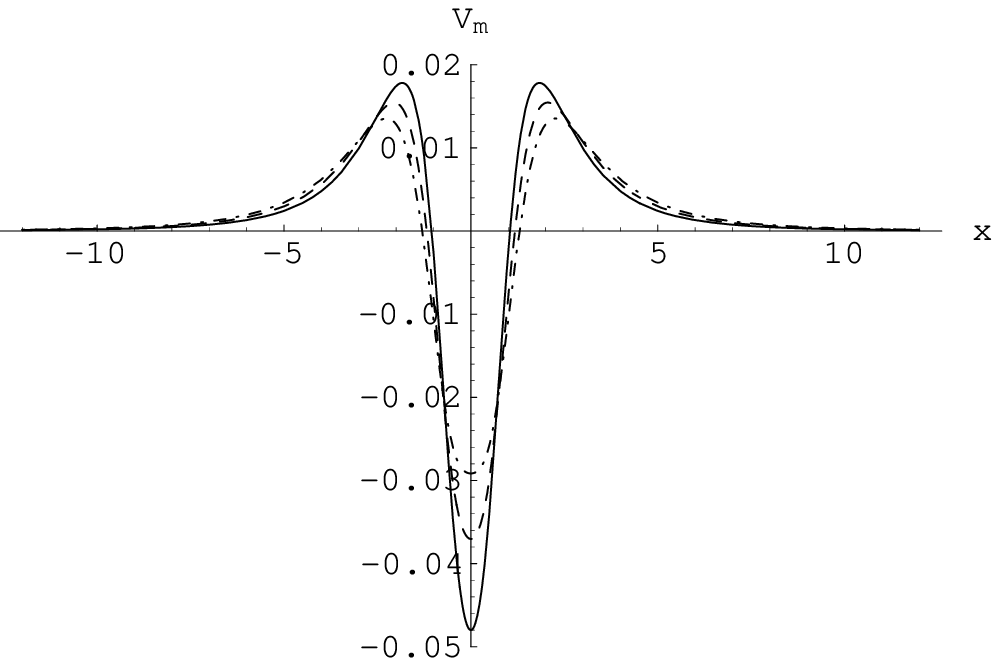}
\end{minipage}}%
\subfigure[]
{
\label{fig:mini:subfig:c}
\begin{minipage}[c]{0.3\textwidth}
\includegraphics[width=5.0cm,height=5.5cm]{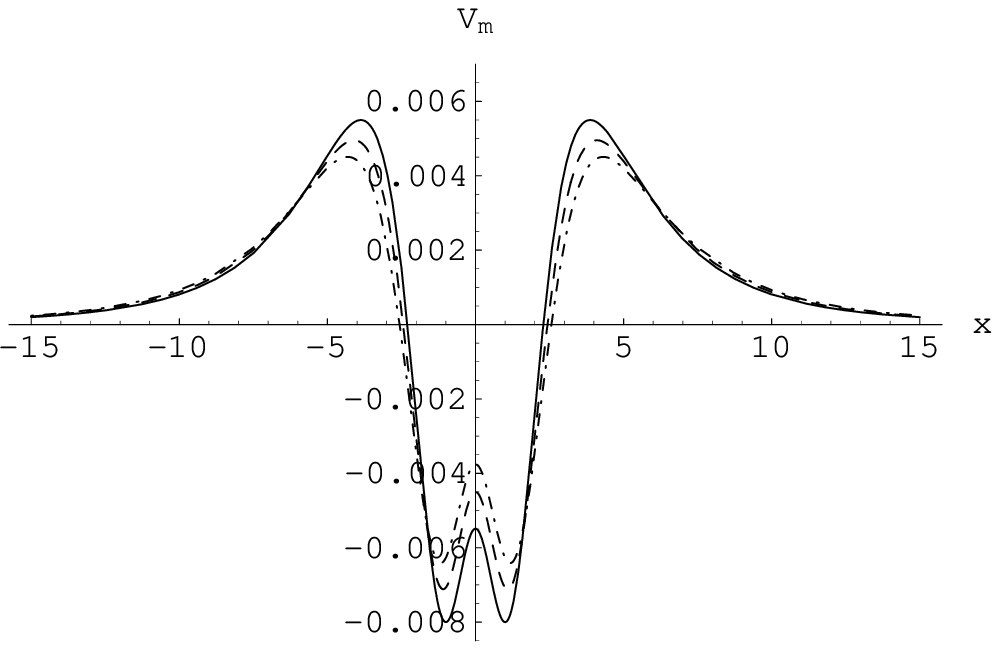}
\end{minipage}}%
\caption{The potential $V_m$ for the mass function (\ref{mf1}) and
the different values of the parameter $b$: (a) $b=0.5$(solid
line), $b=0.6$(dashed line), $b=0.7$(dot-dashed line); (b)
$b=2.5$(solid line), $b=3.0$(dashed line), $b=3.5$(dot-dashed
line);
 (c) $b=9.0$(solid line), $b=10.0$(dashed line), $b=11.0$(dot-dashed line).}
\label{fig:mini:subfig}
\end{figure}

{\bf Example 2:} The effective mass is\cite{gon1,gon2,cai}
\begin{equation}\label{mf2}
m(x) = e^{ - b\left| x \right|},
\end{equation}
where $b \geq 0$ to assure that mass is finite when $x\rightarrow\pm\infty$. Now,
inserting Eq. (\ref{mf2}) into (\ref{eq13}), we get
\begin{equation}
\label{eq81}
 \mu(x) = \left\{
 \begin{aligned}
 & - \frac{2}{b}e^{ - \frac{b}{2}x}, \;\; \; & (b \ne 0,\; \; \; x >
 0)\\
& \frac{2}{b}e^{\frac{b}{2}x}, \; \; \; & (b \ne 0, \; \; \; x <
0)\\
 & x.  \; \; \; & (b = 0)
\end{aligned}
\right.
\end{equation}
The extra interaction due to the mass is
\begin{equation}\label{mfv2}
    V_m=-\frac{3}{32}b^2e^{b\left| x \right|},
\end{equation}
which is monotonously changed as $|x|$ increases from 0.
This $V_m$ has the typical characterization of  Fig.2 for various values of the parameter $b$. It is
a kind of barrier whose width decreases with increasing the parameter $b$.

\begin{figure}[htb]
\includegraphics[width=6.0cm,height=5.5cm]{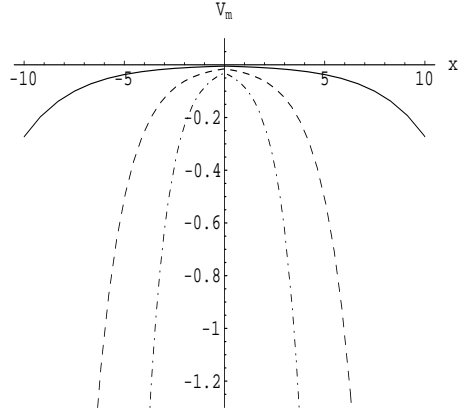}
\caption{The potentials $V_m$
for  the mass function (\ref{mf2}) and the parameters
$b=0.4$(solid line), $b=0.7$(dashed line), $b=1.0$(dot-dashed line),respectively.}
\end{figure}

{\bf Example 3:} The effective mass is of the form\cite{Alh}
\begin{equation}
\label{mf3}
 m(x) = \frac{1}{b + x^2},
\end{equation}
where $b > 0$ to avoid its singularity.
Now, we have
\begin{equation}
\label{eq98} \mu(x) = \ln (x + \sqrt {b + x^2} ).
\end{equation}
The effective potential $V_m$  is
\begin{equation}\label{mfv3}
    V_m=-\frac{2 b + x^2}{8(b + x^2)}.
\end{equation}
This $V_m$ has the behavior of  Fig.3 for various values of the parameter $b$. It is a typically
potential well whose width increases
as the parameter  $b$ is increased.

\begin{figure}[htb]
\includegraphics[width=7.0cm,height=5.0cm]{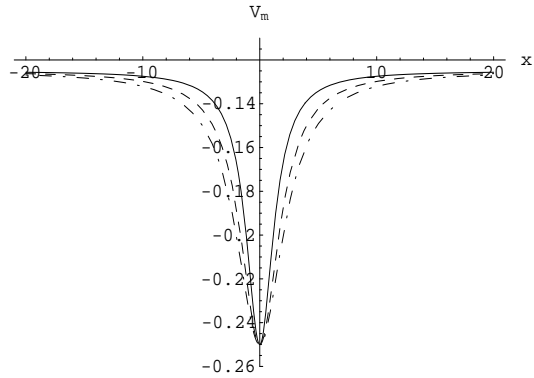}
\caption{The potential $V_m$
for  the mass function (\ref{mf3}) and the parameters $b=2$(solid line), $b=4$(dashed line), $b=6$(dot-dashed line),
respectively.}

\end{figure}

{\bf Example 4:} The effective mass is as follows\cite{dek1, dek2}
\begin{equation}\label{mf4}
m(x) = 1 + \tanh (bx),
\end{equation}
where $a$ is a real parameter. Inserting Eq. (\ref{mf4}) into Eq. (\ref{eq13}), one has
\begin{equation}\label{eq115}
\mu(x) = \left\{
\begin{aligned}
& \frac{\sqrt 2 }{b}\tanh^{-1}\sqrt{\frac{1}{2}\left(1+\tanh(bx)\right)}, \;\;\; &(b\neq0) \\
& x.  \;\;\;&(b=0)
\end{aligned}
\right.
\end{equation}
Similarly, we get the potential  produced from the dependence
of mass on $x$
\begin{equation}\label{mfv4}
    V_m=-\frac{1}{32}b^2 {\rm sech}(a\,x)\,\left[ 7\cosh (b\,x) + \sinh (b\,x) \right]
      \left[ \cosh (2b x) - \sinh (2bx) \right].
\end{equation}
The curves of $V_m$  for various values of the parameter $b$ are displayed  in  Fig.4. It can be seen
that these $V_m$ look like the semi-infinite potential barriers whose widths decrease as the parameter
$b$ increases.

\begin{figure}[htb]
\includegraphics[width=7.0cm,height=4.50cm]{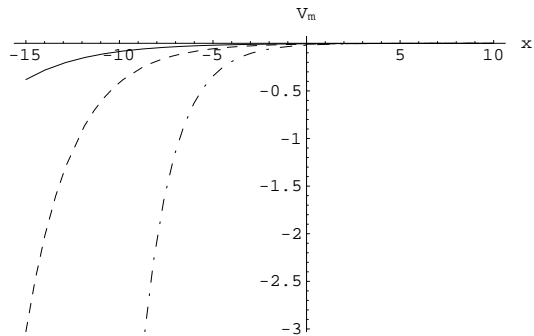}
\caption{The potential $V_m$
for  the mass function (\ref{mf4}) and the parameters $b=0.15$(solid line), $b=0.2$(dashed line),
$b=0.3$(dot-dashed line), respectively.}
\end{figure}

With all above mass functions, we see that the dependence of the mass on the position of the particle will
affect the behaviors of the system through two ways: the argument of the part in the potential that has the same form
as that for the system with constant mass is $\mu(x)$ instead of $x$ and an extra term $V_m$. These effects also
increase the number of the solvable potentials for the same energy eigenvalue.

\section{Remarks and discussions}
In this paper, we use the coordinate transformation method to
study the exact solutions of the PDM Schrodinger equation for several potentials.
The eigenstates of all these systems with PDM can be expressed in terms of three kinds of
special functions. We also give the explicit expressions of the potentials $V_m$
for several mass functions that are used in some physically interested problems and study their properties.
All these results will reduce to
those for the systems with constant mass (see, for example \cite{lev,dutt,lev2}) if we set the mass functions $m(x)$
to be constants in all equations in the above sections.
It should be noted that there are two functions $m(x)$ and $g(x)$ in the PDM case, while only one function
$g(x)$ is concerned in the cases of constant mass. So, the classes with the same form of the energy eigenvalues but
 with different potentials
related by coordinate transformation are enlarged.

In our above discussions, we use the Hamiltonian of the symmetric form (\ref{ham}). If we
adopt the Hamiltonian\cite{von}
\begin{equation}\label{ham2f}
   H=\frac{1}{4}\left(m^{\alpha}\vec{\bf P}m^{\beta}\cdot\vec{\bf P}m^{\gamma}+
   m^{\gamma}\vec{\bf P}m^{\beta}\cdot\vec{\bf P}m^{\alpha}\right)+V(\vec{\bf r}),
\end{equation}
with the condition $\alpha+\beta+\gamma=-1$, then all the results above also hold provided that we replace $V(x)$
by $V_{eff}$ in both Eq. (\ref{ham}) and other related relations, here
\begin{equation}\label{epot}
V_{eff}(x)=V(x)+ \frac{1}{2}\frac{m''}{m^2}-[\alpha(\alpha+\beta+1)+\beta+1]\frac{m'^{2}}{m^3},
\end{equation}
$m'=\frac{dm}{dx}$ and $m''=\frac{d^2 m}{dx^2}$. This is so due to the fact that the
Hamiltonian (\ref{ham2f}) can be rewritten as
\begin{equation}\label{ham21d}
H=-\frac{1}{2}\left[\frac{d}{dx}\frac{1}
          {m(x)}\frac{d}{dx}\right]+V_{eff}(x),
\end{equation}
for the one-dimensional system.

Also, for the three-dimensional systems with
PDM and spherical symmetry, the solution of the system can be written as
the product of angular and radial parts. The radial
Schrodinger equation for the Hamiltonian (\ref{ham}) takes the form
\begin{equation}\label{rsch}
        -\frac{1}{2}\left[\frac{d}{dr}\frac{1}{m}\frac{d}{dr}\right]\phi(r)
        +\left[V(r)+\frac{1}{2m}\frac{l(l+1)}{r^{2}}
        -\frac{m'}{2m^2}\frac{1}{r}\right]\phi(r)=E\phi(r),
\end{equation}
where $m'\equiv\frac{dm(r)}{dr}$, $R(r)=\frac{\phi(r)}{r}$ is the radial
wave function,  $E$ and $l$ are the energy eigenvalue and angular
momentum quantum number of the system, respectively.
Eq. (\ref{rsch}) has the same form with the Schrodinger equation for the Hamiltonian
(\ref{ham21d}), so the results in this paper can also be applied to the  spherically symmetrical systems with PDM
upon some modifications.

We known that the special functions can have some generalizations, such as the q-deformed forms\cite{and},
so the corresponding
differential equations have more general forms than Eq. (\ref{eq5}). The coordinate transformation method
can in principle apply to this generalized case and may give a more general classification to the solvable potentials.

\section{Acknowledgement}
The program is supported by National Natural Science Found for
Outstanding Young Scientists of China under contract 10125521, the
Fund of the Education Ministry  under contract 20010284036, Major
State Basic Research  Development in China under contract
G2000077400, Chinese Academy of Sciences Knowledge Innovation Project(KJCX2-SW-N02),National
Natural Science Found under contract 60371013.

\end{document}